\documentclass[twocolumn,pra,superscriptaddress,showpacs,aps,floatfix]{revtex4-1}
\usepackage{color}
\usepackage{amsmath}
\usepackage{amssymb}
\usepackage{txfonts}
\usepackage{graphicx}
\usepackage{epstopdf}
\usepackage[unicode]{hyperref}
\hypersetup{pdftitle={2D SOC BEC},
 pdfauthor={S.-W. Su,},
 pdfsubject={2D SOC BEC},
  colorlinks=true,
    linkcolor=red,
    citecolor=magenta,
    filecolor=black,
    urlcolor=blue}


\begin{document}

\title{Hidden long-range order in a spin-orbit coupled two-dimensional Bose
gas}

\author{Shih-Wei Su}

\affiliation{Department of Physics and Graduate Institute of Photonics, National
Changhua University of Education, Changhua 50058 Taiwan}

\author{I-Kang Liu}

\affiliation{Department of Physics and Graduate Institute of Photonics, National
Changhua University of Education, Changhua 50058 Taiwan}

\author{Shih-Chuan Gou}
\email{scgou@cc.ncue.edu.tw}

\affiliation{Department of Physics and Graduate Institute of Photonics, National
Changhua University of Education, Changhua 50058 Taiwan}

\affiliation{Physics Division, National Center for Theoretical Sciences, Hsinchu
30013, Taiwan}

\author{Renyuan Liao}

\affiliation{College of Physics and Energy, Fujian Normal University, Fuzhou 350108,
China}

\author{Oleksandr Fialko}
\affiliation{Dodd-Walls Centre for Photonics and Quantum Technology and Centre
for Theoretical Chemistry and Physics, Institute for Natural and Mathematical
Sciences, Massey University, Auckland, New Zealand}

\author{Joachim Brand}
\email{J.Brand@massey.ac.nz}
\affiliation{Dodd-Walls Centre for Photonics and Quantum Technology and Centre
for Theoretical Chemistry and Physics, New Zealand Institute for Advanced
Study, Massey University, Auckland, New Zealand}

\begin{abstract}
A spin-orbit coupled two-dimensional (2D) Bose gas is shown to simultaneously
possess quasi and true long-range order in the total and relative
phase sectors, respectively. The total phase undergoes a  Berenzinskii-Kosterlitz-Thouless
transition to a low temperature phase with quasi long-range order, as expected for a two-dimensional quantum gas. Additionally,
the relative phase undergoes an Ising-type transition building up
true long-range order, which is induced by the anisotropic spin-orbit
coupling. Based on the Bogoliubov approach, expressions for the total-
and relative-phase fluctuations are derived analytically for the low
temperature regime. Numerical simulations of the stochastic projected
Gross-Pitaevskii equation (SPGPE) give a good agreement with the analytical
predictions. 
\end{abstract}
\maketitle
\section{introduction}

Spatial dimensionality and interactions play crucial roles in the
physics of phase transitions. The governing  paradigm is the Hohenberg-Mermin-Wagner
theorem~\cite{Mermin1966,Hohenberg1967}, which asserts that a uniform
infinite system with short-range interaction possessing continuous
symmetries cannot exhibit long-range order (LRO) at finite temperatures
in $d\leq2$ dimensions. In the context of Bose gases, this implies
the nonexistence of Bose-Einstein condensation (BEC) in dimension
$d\leq2$ in the thermodynamic limit. Instead, a 2D Bose gas can develop a 
quasi LRO in the low-temperature
phase, characterized by an algebraically decaying correlation function,
and undergoes a phase transition to the high-temperature phase, where
the correlation between particles decays exponentially. This mechanism
is known as the Berezinskii-Kosterlitz-Thouless (BKT) transition~\cite{Berezinskii1971,Kosterlitz1973,Hadzibabic2006,Foster2010}.

Recent advances in the manipulation of ultra-cold atoms have made it possible to study uniform 2D quantum degenerate gases  \cite{Gaunt2013,Chomaz2015} and thus it is timely to probe the unexplored aspects of two-dimensional phase transitions.
To this end, we are particularly
interested in the condensation of spin-orbit coupled pseudo spin-1/2
Bose gases~\cite{Lin2011}, which have attracted a great deal of attention
in recent years~\cite{Campbell2011,Xu2011,Sinha2011,Su2012,Hu2012,Zhou2013a,Cui2013,Jian2011,Li2012,Li2013,Liao2014,Sun2015,Su2015,Su2016}.
The spin-orbit coupling (SOC) here refers to a synthetic gauge field
originating from the laser-assisted coupling between the atomic center-of-mass
motion and the internal degrees of freedom~\cite{Lin2011,Campbell2011,Dalibard2011}.
Synthetic SOC in ultra-cold gases has so far been realized in one-dimensional (1D)
\cite{Lin2011} and 2D form \cite{Huang2016,Wu2016}
and has become a tunable resource \cite{Jimenez-Garcia2015}, with
more exotic realizations proposed \cite{Anderson2012,Anderson2013}.
For an ideal two-component Bose gas, the presence of SOC can enhance
the density of states at low energies, making the system more susceptible
to both quantum and thermal fluctuations and thus preventing the atoms
from condensing~\cite{Hu2012,Zhou2013a,Cui2013}. On the other hand, 
the interatomic interactions can stabilize the condensate, and enhanced
condensation due to SOC was found in superfluid Fermi gases \cite{Zhou2012,He2012b,Xu2015}.
It is thus anticipated that the competition between fluctuations and
interactions in the presence of SOC can drastically affect the mechanism
of the BEC phase-transition. 
Recently the thermal properties of
spin-orbit coupled 2D Bose gases have been investigated and 
extended scenarios of BKT physics reaching from relative suppression of superfluidity to fractionalised vortex phases
have been predicted~\cite{Jian2011,Liao2014}. In these studies,
the corresponding effective theories were derived in terms of
the total-phase degree of freedom by integrating out the relative-phase
counterpart. Since the variables representing respectively the total-
and relative-phase sectors are entwined via SOC,
a more complete picture of the nature of the superfluid phase transition can be obtained
by considering all degrees of freedom. The aim of the current work is to address this issue.

In this paper we study the low-temperature properties of a spin-orbit coupled two-dimensional Bose gas in the plane-wave phase with Bogoliubov theory and simulations with the stochastic projected Gross-Pitaevskii equation (SPGPE)~\cite{Blakie2008,Rooney2012,Su2013,Bradley2014}. We find that quasi-long-range order in the total phase of the pseudo-spin-$\frac{1}{2}$ superfluid coexists with true long range order of the relative phase between the two spin components. 

The organization of this paper is as follows. In Section II, the exact solutions to Bogoliubov-de Gennes equations pertinent to the low-lying excitations of a 2D Bose gas with anisotropic SOC are presented, which reveal the low-temperature properties of the gas. In Section III, two-point correlation functions of the total- and relative phases are calculated both analytically and numerically.  Based on the analytical results in Section II, close-form expressions of the phase correlation functions are derived. Meanwhile, to explore the essence of BEC phase transition in the current system on an ab initio basis, we perform  SPGPE simulations to evaluate the correlation functions over a wide range of temperatures. The attributes of phase transitions in the total- and relative phases are verified according to the behavior of the correlation functions and the underlying physics is addressed. Concluding remarks are given in Section IV, including a discussion on the experimental implementation for measuring the hidden LRO of the system. Finally, auxiliary calculations and derivations are placed in Appendix.

\section{Formulae}

The system under study is described by the Hamiltonian 

\begin{eqnarray}
\hat{H} & = & \int d^{2}r\,\left[\mathbf{\hat{\Psi}}^{\dagger}\hat{H}_{\mathrm{sp}}\mathbf{\hat{\mathbf{\Psi}}}+\frac{g_{11}}{2}(\hat{\Psi}_{1}^{\dagger}\hat{\Psi}_{1})^{2}+\frac{g_{22}}{2}(\hat{\Psi}_{2}^{\dagger}\hat{\Psi}_{2})^{2}\right.\nonumber \\
 &  & \left.+g_{12}\hat{\Psi}_{1}^{\dagger}\hat{\Psi}_{1}\hat{\Psi}_{2}^{\dagger}\hat{\Psi}_{2}\right],\label{eq:Hamiltonian-SOC}
\end{eqnarray}
where $\mathbf{\hat{\Psi}}=(\hat{\Psi}_{1},\hat{\Psi}_{2})^{\mathrm{T}}$
is the two-component spinor field operator and $\hat{H}_{\mathrm{sp}}=-\hbar^{2}\nabla^{2}/2m+\kappa_{x}\hat{p}_{x}\hat{\sigma}_{x}+\kappa_{y}\hat{p}_{y}\hat{\sigma}_{y}$
is the single-particle Hamiltonian with $\kappa_{x,y}$ the spin-orbit
coupling strengths along different directions and $\hat{\sigma}_{x,y}$ the Pauli matrices. The
inter- and intra-species atomic interaction strengths are characterized
by $g_{12}$ and $g_{ii}$ $(i=1,2)$, respectively. Without loss
of generality, we shall assume that the intra-species interactions
are identical, i.e., $g_{11}=g_{22}\equiv g$. Diagonalizing the single-particle
Hamiltonian yields two dispersion branches, $\epsilon_{\pm}=p^{2}/2m\pm(\kappa_{x}^{2}p_{x}^{2}+\kappa_{y}^{2}p_{y}^{2})^{1/2}$,
and the corresponding eigenvectors, $\phi_{\mathbf{k}}^{\pm}=(1,\pm e^{i\varphi_{\mathbf{k}}})^{\mathrm{T}}e^{i\mathbf{p}\cdot\mathbf{x}/\hbar}/\sqrt{2}$,
where $\varphi_{\mathbf{k}}=\mathrm{arg}(\kappa_{x}p_{x}+i\kappa_{y}p_{y})$
\cite{Wang2010}. For anisotropic SOC ($\kappa_{x}\neq\kappa_{y}$)
the single-particle ground state lies in the lower branch, and is
two-fold degenerate at $\mathbf{k}=\pm m\kappa_{x}\mathbf{e}_{x}$
($\pm m\kappa_{y}\mathbf{e}_{y}$) for $\left|\kappa_{x}\right|>\left|\kappa_{y}\right|$
($|\kappa_{x}|<|\kappa_{y}|$). On the other hand the single-particle
ground state is infinitely degenerate on the Rashba ring of radius
$|\mathbf{p}|=m\kappa$ in momentum space for isotropic SOC ($|\kappa_{x}|=|\kappa_{y}|\equiv\kappa$).

For an interacting gas, depending on the interatomic interaction strengths,
the ground state phases are characterized by the plane waves
corresponding to the minima of the single-particle dispersion. For $g>g_{12}$,
the ground state is a single plane-wave (PW) state while for $g<g_{12}$
the ground state is a standing-wave created by the superposition of
two plane waves carrying opposite momenta~\cite{Wang2010}. In the
following calculation, we shall work in the dimesionless units where the
length, time, and energy are scaled by $a_{h}=\sqrt{\hbar/m\omega_{0}}$,
$1/\omega_{0}$, and $\hbar\omega_{0}$, respectively, with $m$ the atomic
mass and $\omega_{0}$ the transverse trapping frequency. In
the following, the dimensionless interatomic interaction strengths
and SOC strengths are denoted by $\tilde{g}_{ij}$ and $\tilde{\kappa}_{x,y}$,
respectively.

Within the framework of mean-field theory, the dynamics of Bose gases is determined by the Gross-Pitaevskii energy functional
$\mathcal{E}[\mathbf{\Psi}^{*},\mathbf{\Psi}]=\hat{\left\langle H\right\rangle }$,
where the Bose fields in Eq.~(\ref{eq:Hamiltonian-SOC}) are replaced
by the complex classical-field wave functions, $\Psi_{j}=\langle\hat{\Psi}_{j}\rangle$.
The Gross-Pitaevskii equation, $i\hbar\partial_{t}\Psi_{j}=\mathcal{L}_{j}\Psi_{j}$,
can be derived via the Hartree variational principle {(}see Appendix{)}. For
definiteness and to assure the validity of the mean-field approach,
we will consider anisotropic SOC and focus on the PW state in what
follows, which avoids the degeneracies and ambiguities of scenarios
with higher symmetry \cite{Zhou2013a}. At zero temperature, the PW
state wave function is $\mathbf{\mathbf{\Psi}}^{0}=(\Psi_{1}^{0},\Psi_{2}^{0})^{\mathrm{T}}=\sqrt{n}e^{-i\tilde{\kappa}_{x}x}(1,1)^{\mathrm{T}}$
where we assume that the condensation occurs at $\mathbf{p}=(-|\tilde{\kappa}_{x}|,0)$
and $n$ is the total particle density. Furthermore, the PW state
is characterized by a non-vanishing pseudo spin density, $\mathbf{S}=\sum_{\alpha,\beta}\Psi_{\alpha}^{*}\boldsymbol{\sigma}_{\alpha\beta}\Psi_{\beta}$,
along $x$ direction, $\mathbf{S}^{0}=n\mathbf{e}_{x}$. To investigate
the low-lying excitations, we adopt the Bogoliubov formulation where
the total wave function is decomposed as $\Psi_{j}=e^{-i\mu t}e^{-i\tilde{\kappa}_{x}x}(\Psi_{j}^{0}+\delta\Psi_{j})$
with $\mu$ the chemical potential and $\delta\Psi_{j}$ the low-lying
excitation. Inserting $\delta\Psi_{j}=\sum_{\mathbf{q}}(u_{j}^{\mathbf{q}}e^{i(\mathbf{q}\cdot\mathbf{r}-\omega t)}-v_{j}^{\mathbf{q}*}e^{-i(\mathbf{q}\cdot\mathbf{r}-\omega t)})/\sqrt{A}$,
where $A$ is the area of system and $\omega$ is the excitation energy
of the mode with momentum $\mathbf{q}$, into the Gross-Pitaevskii
equation yields the Bogoliubov-de Gennes equation {(}also see Appendix{)} 
\begin{widetext}
\begin{equation}
\left(\begin{array}{cccc}
\mathcal{L}_{0}-\tilde{\kappa}_{x}q_{x} & -\tilde{g}n & \tilde{g}_{12}n+h_{soc}-\tilde{\kappa}_{x}^{2} & -\tilde{g}_{12}n\\
\tilde{g}n & -\mathcal{L}_{0}-\tilde{\kappa}_{x}q_{x} & \tilde{g}_{12}n & h_{soc}^{*}+\tilde{\kappa}_{x}^{2}-\tilde{g}_{12}n\\
\tilde{g}_{12}n+h_{soc}^{*}-\tilde{\kappa}_{x}^{2} & -\tilde{g}_{12}n & \mathcal{L}_{0}-\tilde{\kappa}_{x}q_{x} & -\tilde{g}n\\
\tilde{g}_{12}n & h_{soc}+\tilde{\kappa}_{x}^{2}-\tilde{g}_{12}n & \tilde{g}n & -\mathcal{L}_{0}-\tilde{\kappa}_{x}q_{x}
\end{array}\right)\left(\begin{array}{c}
u_{1}^{\mathbf{q}}\\
v_{1}^{\mathbf{q}}\\
u_{2}^{\mathbf{q}}\\
v_{2}^{\mathbf{q}}
\end{array}\right)=\omega\left(\begin{array}{c}
u_{1}^{\mathbf{q}}\\
v_{1}^{\mathbf{q}}\\
u_{2}^{\mathbf{q}}\\
v_{2}^{\mathbf{q}}
\end{array}\right),\label{eq:BdG}
\end{equation}
where $\mathcal{L}_{0}=q^{2}/2+\tilde{g}n+\tilde{\kappa}_{x}^{2}$,
$h_{soc}=\tilde{\kappa}_{x}q_{x}-i\tilde{\kappa}_{y}q_{y}$ and $u_{j}^{\mathbf{q}},\,v_{j}^{\mathbf{q}}$
satisfy the normalization condition $\sum_{j}|u_{j}^{\mathbf{q}}|^{2}-|v_{j}^{\mathbf{q}}|^{2}=1$.
For the fully anisotropic SOC ($\tilde{\kappa}_{y}=0$), Eq.~(\ref{eq:BdG})
is solved with the two distinct energy dispersion relations of the
excitation: 
\end{widetext}

\begin{equation}
\omega_{\mathrm{t}}^{\mathbf{q}}=\sqrt{(\xi_{\mathrm{t}}^{\mathbf{q}})^{2}-(\tilde{g}+\tilde{g}_{12})^{2}n^{2}},\label{eq:excitation-total}
\end{equation}
with $\xi_{\mathrm{t}}^{\mathbf{q}}=q^{2}/2+\left(\tilde{g}+\tilde{g}_{12}\right)n$
and the eigenvector $\delta\mathbf{\Psi}_{\mathrm{t}}^{\mathbf{q}}\sim(u_{\mathrm{t}}^{\mathbf{q}},v_{\mathrm{t}}^{\mathbf{q}},u_{\mathrm{t}}^{\mathbf{q}},v_{\mathrm{t}}^{\mathbf{q}})^{\mathrm{T}}$; 
\begin{equation}
\omega_{\mathrm{r}}^{\mathbf{q}}=-2q_{x}\tilde{\kappa}_{x}+\sqrt{(\xi_{\mathrm{r}}^{\mathbf{q}})^{2}-(\tilde{g}-\tilde{g}_{12})^{2}n^{2}}\label{eq:excitation-relative}
\end{equation}
with $\xi_{\mathrm{r}}^{\mathbf{q}}=q^{2}/2+\left(\tilde{g}-\tilde{g}_{12}\right)n+2\tilde{\kappa}_{x}^{2}$
and the eigenvector $\delta\mathbf{\Psi}_{\mathrm{r}}^{\mathbf{q}}\sim(u_{\mathrm{r}}^{\mathbf{q}},v_{\mathrm{r}}^{\mathbf{q}},-u_{\mathrm{r}}^{\mathbf{q}},-v_{\mathrm{r}}^{\mathbf{q}})^{\mathrm{T}}$.

Equation~(\ref{eq:excitation-total}) represents a gapless mode corresponding
to the total-phase excitation that is immune to SOC. On the other hand,
Eq.~(\ref{eq:excitation-relative}) indicates a mode corresponding
to the relative-phase spin excitation where the effect of SOC acts to
open a gap but also shift the minimum of the dispersion. For non-vanishing
$\kappa_{y}$ the eigenenergies and eigenvectors can be calculated
numerically and the above conclusion remains valid.

\section{results and discussions}

\begin{figure}
\includegraphics[width=1\columnwidth]{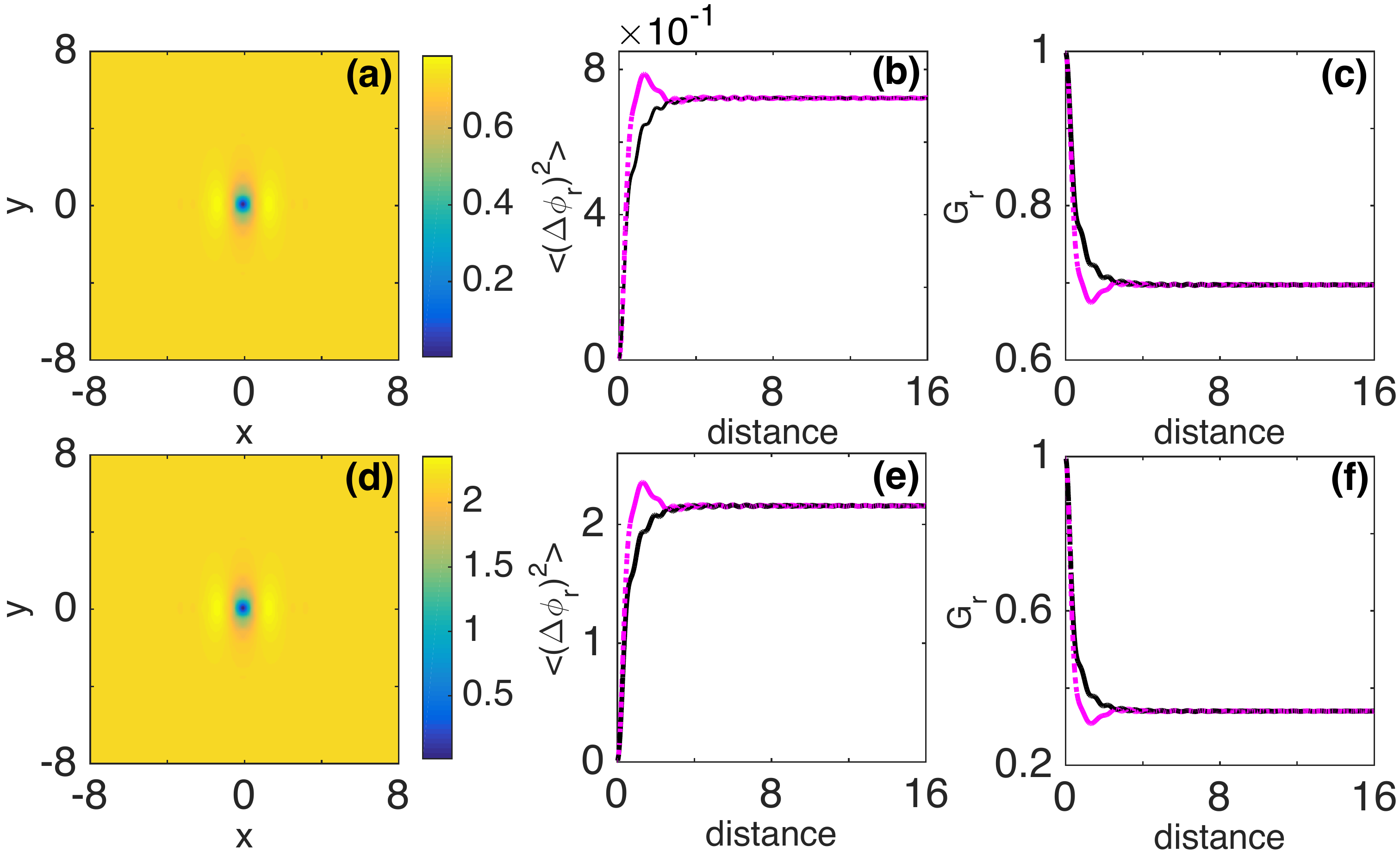}\caption{(Color online). Relative-phase fluctuations from Bogoliubov theory
at two different temperatures. Panels (a) and (d) show the relative-phase
fluctuations $\left\langle (\Delta\phi_{\mathrm{r}})^{2}\right\rangle $
from Eq.~\eqref{eq:fluctuations-int} where the axes denote the separations
$x=x'-x''$ and $y=y'-y''$. A plateau is seen to develop at large
separation. The fluctuations $\left\langle (\Delta\phi_{\mathrm{r}})^{2}\right\rangle $
are also shown in panels (b) and (e) along the $x$- (magenta) and
the $y$-axis (black) while panels (c) and (f) depict the correlation
function $G_{\mathrm{r}}$ from Eq.~\eqref{eq:correlation} with
the same color coding. The temperature is set to $T/T_{\mathrm{BKT}}^{\infty}\approx0.44$
for panels (a), (b), (c) and to $T/T_{\mathrm{BKT}}^{\infty}\approx1.33$
for panels (d), (e), (f), and $\mu=13$, $\tilde{g}_{12}/\tilde{g}=0.9$
and $(\tilde{\kappa}_{x},\tilde{\kappa}_{y})=(1,0)$.}
\label{Fig1} 
\end{figure}

To study the phase fluctuations in the spin-orbit coupled Bose gas,
the Bose field can be expressed as~\cite{Pethick2008} 
\begin{eqnarray}
\mathbf{\hat{\Psi}}=\left(\begin{array}{c}
\hat{\Psi}_{1}(\mathbf{r}')\\
\hat{\Psi}_{2}(\mathbf{r}')
\end{array}\right) & = & \sqrt{n}e^{i\hat{\phi}_{\mathrm{t}}(\mathbf{r}')}\left(\begin{array}{c}
e^{i\hat{\phi}_{\mathrm{r}}(\mathbf{r}')}\\
e^{-i\hat{\phi}_{\mathrm{r}}(\mathbf{r}')}
\end{array}\right)\,,\label{eq:fluctuations}
\end{eqnarray}
where $\hat{\phi}_{\mathrm{t,r}}$ denote the total- and relative-phase
operators, respectively and we have neglected the density fluctuations.
For small fluctuations, Eq.~(\ref{eq:fluctuations}) can be expanded
to the first order which gives $\hat{\phi}_{\mathrm{t,r}}=\sum_{\mathbf{q}}[(\mathcal{U}_{\mathrm{t,r}}^{\mathbf{q}}+\mathcal{V}_{\mathrm{t,r}}^{\mathbf{q}})\hat{\alpha}_{\mathrm{t,r}}^{\mathbf{q}}-h.c.]/2i\sqrt{n},$
where $\hat{\alpha}_{\mathrm{t,r}}^{\mathbf{q}}$ ($\hat{\alpha}_{\mathrm{t,r}}^{\mathbf{q}\dagger}$)
is the annihilation (creation) operator that destroys (creates) the
excitation in the corresponding branch $\omega_{\mathrm{t,r}}^{\mathbf{q}}$
and $(\mathcal{U}_{\mathrm{t,r}}^{\mathbf{q}},\mathcal{V}_{\mathrm{t,r}}^{\mathbf{q}})=(u_{\mathrm{t,r}}^{\mathbf{q}},v_{\mathrm{t,r}}^{\mathbf{q}})e^{i\mathbf{q}\cdot\mathbf{r}}/\sqrt{A}$
is the amplitude of Bogoliubov excitation. In the linear approximation,
the total- and relative-phase operators are decoupled and can be expressed
in terms of the excitations $\delta\mathbf{\Psi}_{\mathrm{t}}^{\mathbf{q}}$
and $\delta\mathbf{\Psi}_{\mathrm{r}}^{\mathbf{q}}$, respectively.
The two-point phase correlation functions are given by 
\begin{equation}
G_{\mathrm{t,r}}\left(\mathbf{r}',\mathbf{r}''\right)=\left\langle e^{i\hat{\phi}_{\mathrm{t,r}}(\mathbf{r}')-i\hat{\phi}_{\mathrm{t,r}}(\mathbf{r''})}\right\rangle =e^{-\left\langle (\Delta\phi_{\mathrm{t,r}})^{2}\right\rangle /2},\label{eq:correlation}
\end{equation}
where $\left\langle \cdots\right\rangle $ denotes the ensemble average
and $\Delta\phi_{\mathrm{t,r}}=\hat{\phi}_{\mathrm{t,r}}(\mathbf{r}')-\hat{\phi}_{\mathrm{t,r}}(\mathbf{r}'')$.
The thermal average can be expressed in terms of the Bogoliubov amplitudes
\begin{equation}
\left\langle (\Delta\phi_{\mathrm{t,r}})^{2}\right\rangle =\int\frac{d^{2}q}{\pi n}(N_{\mathrm{t,r}}^{\mathbf{q}}+\frac{1}{2})(u_{\mathrm{t,r}}^{\mathbf{q}}+v_{\mathrm{t,r}}^{\mathbf{q}})^{2}\text{\ensuremath{\sin}}^{2}\frac{\mathbf{q}\cdot\mathbf{r}}{2},\label{eq:fluctuations-int}
\end{equation}
where $N_{\mathrm{t,r}}^{\mathbf{q}}=1/[\exp(\omega_{\mathrm{t,r}}^{\mathbf{q}}/T)-1]$
is the Bose-Einstein distribution function with $T$ the temperature
measured in units of $\hbar\omega_{0}/k_{B}$. Due to translational
invariance the averaged phase fluctuations and the correlation function
only depend on the separation $\left|\mathbf{r}\right|=\left|\mathbf{r}'-\mathbf{r}''\right|$.
The Bogoliubov amplitudes in the integrand are $(u_{\mathrm{t}}^{\mathbf{q}}+v_{\mathrm{t}}^{\mathbf{q}})^{2}=[\xi_{\mathrm{t}}^{\mathbf{q}}+(\tilde{g}+\tilde{g}_{12})n]/2\omega_{\mathrm{t}}^{\mathbf{q}}$
and $(u_{\mathrm{r}}^{\mathbf{q}}+v_{\mathrm{r}}^{\mathbf{q}})^{2}=[\xi_{\mathrm{r}}^{\mathbf{q}}+(\tilde{g}-\tilde{g}_{12})n]/2(\omega_{\mathrm{r}}^{\mathbf{q}}+2\tilde{\kappa}_{x}q_{x})$. The total-phase fluctuation shown in Eq.~(\ref{eq:fluctuations-int})
exhibits an infrared divergence similar to that of a 
2D scalar Bose gas. Accordingly, the total-phase correlation function is shown in Fig.~\ref{FigA1} in Appendix. In the thermodynamic limit it is expected that
the long-range correlation $\lim_{|\mathbf{r}|\rightarrow\infty}e^{-\left\langle (\Delta\phi_{\mathrm{r}})^{2}\right\rangle /2}$
would be destroyed by the total-phase fluctuations, leading to the
BKT-type physics which is characterized by the quasi LRO
as discussed in Ref.~\cite{Jian2011}. The BKT transition temperature
for the 2D scalar Bose gas is given by $T_{\mathrm{BKT,scalar}}^{\infty}=2\pi\hbar^{2}n/\{mk_{B}\ln[(380\pm3)/\tilde{g}_{0}]\}$
with $\tilde{g}_{0}$ the dimensionless interaction strength \cite{Prokofev2001,Prokofev2002}.
Comparing the excitation spectrum of the 2D scalar Bose gas with the
in-phase excitation energy $\omega_{\mathrm{t}}^{\mathbf{q}}$, the
BKT transition temperature $T_{\mathrm{BKT}}^{\infty}$ for the total-phase
degree of freedom can be estimated by replacing $\tilde{g}_{0}$ with
$\tilde{g}+\tilde{g}_{12}$. On the contrary the fluctuation $\left\langle (\Delta\phi_{\mathrm{r}})^{2}\right\rangle $
is suppressed due to the gapped and anisotropic excitation energy,
leading to the existence of true LRO in the relative-phase
correlation. The relative-phase fluctuations evaluated from Eq.~\eqref{eq:fluctuations-int}
are shown in Fig.~\ref{Fig1}. The plateau at a constant value for
a separation $\left|\mathbf{r}\right|$ larger than $\approx4=4\tilde{\kappa}_{x}^{-1}\approx20\xi$,
where $\xi=1/\sqrt{2\mu}$ is the zero-temperature healing length
in scaled units. It is remarkable that the length scale for plateau
formation is independent of temperature while the magnitude decreases
with increasing temperature. Additionally, the effect of anisotropic
SOC appears in the spatial variation at short length scales as clearly
seen in Fig.~\ref{Fig1}.

To verify the analytical prediction, we numerically calculate the
first-order correlation functions by evolving the stochastic projected
Gross-Pitaevskii equation~\cite{Blakie2008,Rooney2012,Su2013,Bradley2014}
\begin{eqnarray}
d\Psi_{j} & = & \mathcal{P}\left\{ -i\mathcal{L}_{j}\Psi_{j}dt+\Gamma(\mu-\mathcal{L}_{j})\Psi_{j}dt+dW_{j}\right\} ,\label{eq:SPGPE}
\end{eqnarray}
where $\mathcal{P}$ is the projection operator restricting the evolution
to the region of $E\,$$\,<\,$$\,\epsilon_{\mathrm{cut}}$, $\mu$
the chemical potential, $\Gamma$ the growth rate and $dW_{j}$ is
the complex white noise satisfying the fluctuation-dissipation relation
$\left\langle dW_{j}^{*}\left(\mathbf{r}',t\right)dW_{k}\left(\mathbf{r}'',t\right)\right\rangle =2\Gamma T\delta\left(\mathbf{r}',\mathbf{r}''\right)\delta_{jk}dt$.
The phase correlation function of Eq.~(\ref{eq:correlation}) can
be numerically computed via the expression $G_{\mathrm{t,\,r}}(\mathbf{r}',\mathbf{r}'')=\frac{1}{N_{s}}\sum_{j=1}^{N_{s}}\exp[i\phi_{\mathrm{t,r}}(\mathbf{r}',t_{j})-i\phi_{\mathrm{t,r}}(\mathbf{r}'',t_{j})],$
where ${t_{j}}$ is a set of $N_{s}$ times at which the field is
sampled after the system reaches equilibrium \cite{Blakie2008,Foster2010}.
In the numerical simulation, we consider the parameters $\mu=13$,
$\epsilon_{\mathrm{cut}}\thickapprox42$, $\tilde{g}_{12}/\tilde{g}=0.9$
and $(\tilde{\kappa}_{x},\tilde{\kappa}_{y})=(1,0)$ at various temperatures.
To obtain an equilibrated sample for calculating the correlation function,
we let the system evolve for a sufficiently long time ($\gg1/\Gamma$)
and then take $10^{3}$ samples to implement the averaging.

\begin{figure}
\includegraphics[width=0.95\columnwidth]{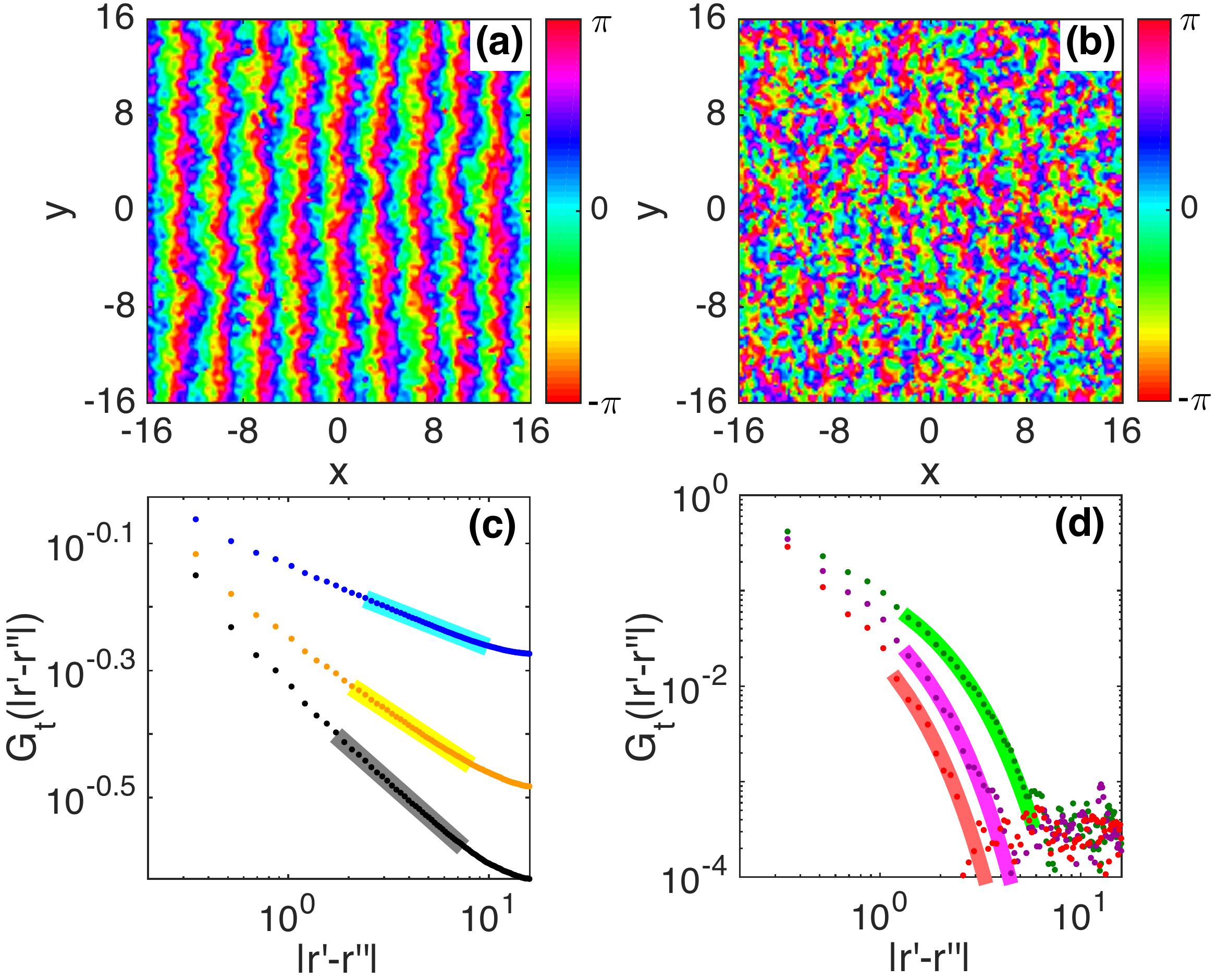}\caption{(Color online). Total phase from stochastic simulations of Eq.~\eqref{eq:SPGPE}.
Panels (a) and (b) depict the snapshot of the total-phase profile $\mathrm{arg}(\Psi_{1})+\mathrm{arg}(\Psi_{2})$
at $T/T_{\mathrm{BKT}}^{\infty}\approx0.44$ and $1.33$ respectively.
The correlation function $G_{\mathrm{t}}(|\mathbf{r'}-\mathbf{r}''|)$
is shown on a doubly-logarithmic scale in panels (c) and (d). Dots
represent numerical data and solid lines are algebraic fits for the
lower temperatures in panel (c) and exponential fits in panel (d).
The temperatures are $T/T_{\mathrm{BKT}}^{\infty}\approx0.44$ (blue),
$0.67$ (orange), $0.78$ (black), $1.33$ (green), $1.56$ (magenta)
and $1.78$ (red). }
\label{Fig2} 
\end{figure}

Figure~\ref{Fig2} depicts the total-phase profile and correlation
at various temperatures. At low temperatures the total-phase exhibits
the periodic structure shown in Fig.~\ref{Fig2}(a), a consequence
of the PW state entailing the phase factor $e^{-2i\tilde{\kappa}_{x}x}$.
At high temperatures, the increasing thermal fluctuations smear out
the quasi periodic structure in Fig.~\ref{Fig2}(a) and results in
a fluctuating total-phase profile as shown in Fig.~\ref{Fig2}(b).
Further analyses of the total-phase correlation are shown in Figs.~\ref{Fig2}(c)
and~\ref{Fig2}(d). For $T<T_{\mathrm{BKT}}^{\infty}$, the results
are consistent with algebraic decay of the correlation function while
for $T>T_{\mathrm{BKT}}^{\infty}$ the correlation function decays
exponentially, a defining feature of the BKT transition.

\begin{figure}[ht]
\includegraphics[width=0.9\columnwidth]{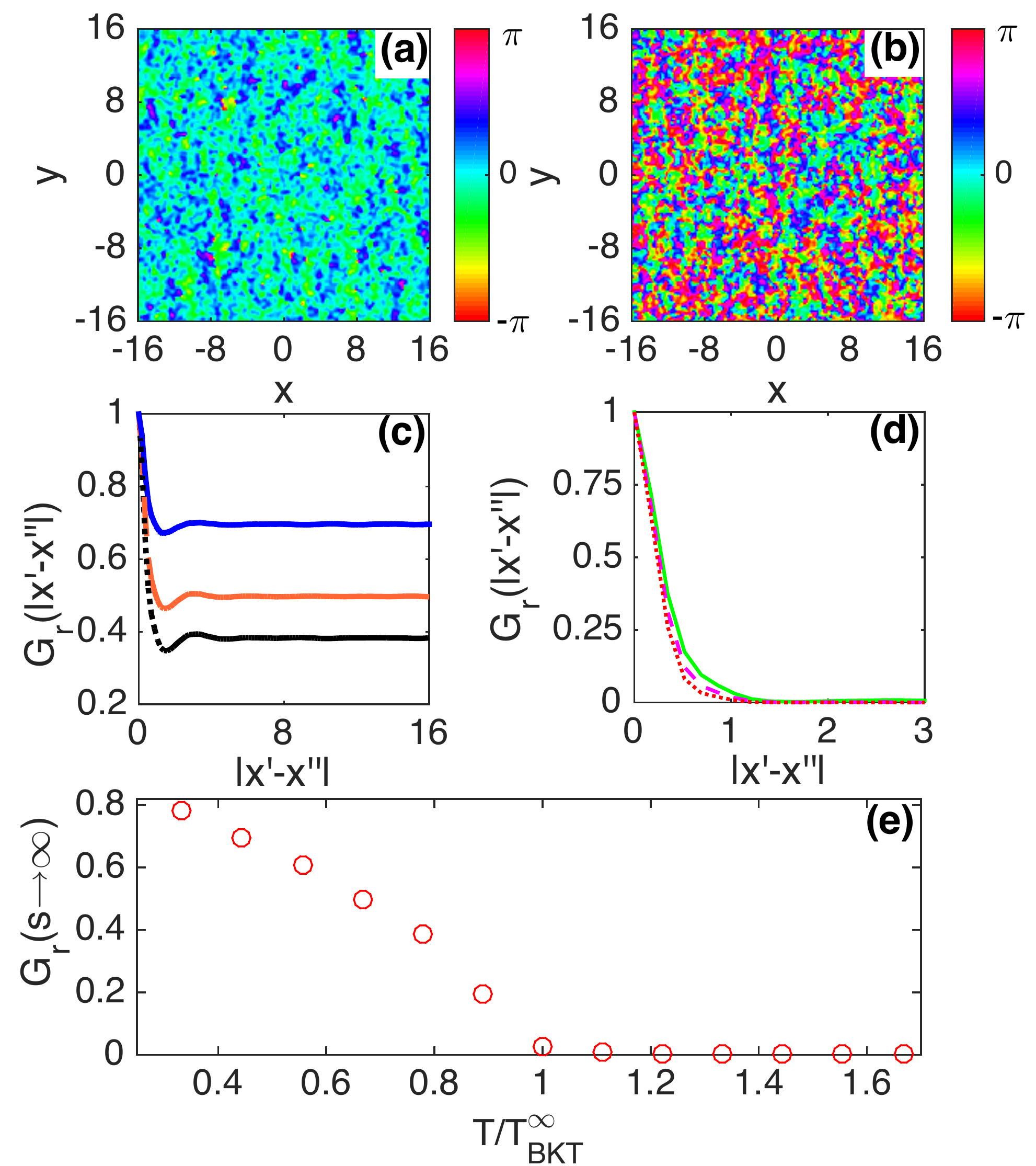}\caption{(Color online). Relative phase from stochastic simulations. Panels
(a) and (b) depict the snapshots of the relative-phase profile $\mathrm{arg}(\Psi_{1})-\mathrm{arg}(\Psi_{2})$
at $T/T_{\mathrm{BKT}}^{\infty}\approx0.44$ and $1.33$ respectively.
The correlation function $G_{\mathrm{r}}$ is plotted in panels (c)
and (d) along the SOC direction. While plateaus are reached in (c)
for $T/T_{\mathrm{BKT}}^{\infty}\approx0.44$, (blue solid line),
$0.67$ (orange dashed line), $0.78$ (black dotted line), the correlation
function quickly decays to zero for the higher temperatures $T/T_{\mathrm{BKT}}^{\infty}\approx1.33$(green
solid line), $1.56$ (magenta dashed line) and $1.78$ (red dotted
line) in panel (d). Panel (e) shows the plateau values for the phase
correlation $G_{\mathrm{r}}(\left|x'-x''\right|\rightarrow\infty)$
vs temperature.}
\label{Fig3} 
\end{figure}

The relative-phase profiles and the correlation functions are shown
in Fig.~\ref{Fig3}. Unlike the total-phase case, thermal fluctuations
in relative-phase sector are suppressed in the low-temperature regime,
as shown in Fig.~\ref{Fig3}(a), and the corresponding correlation
function shown in Fig.~\ref{Fig3}(c) develops a plateau structure
at large separation, implying an established LRO.
On the other hand, the strong thermal fluctuations in the high-temperature
regime completely randomize the phase distribution, leading to an
exponentially decaying correlation function, as shown in Fig.~\ref{Fig3}(d).
The value of phase correlation decreases with increasing
temperature and eventually vanishes for $T>T_{\mathrm{BKT}}^{\infty}$,
as shown in Fig.~\ref{Fig3}(d) and~\ref{Fig3}(e). We note that
in Fig.~\ref{Fig3}(c) the correlation function exhibits oscillations
at small separation along $x$ direction. This qualitatively agrees
with the oscillations in Fig.~\ref{Fig1}(c) and~\ref{Fig1}(f),
which can be attributed to the anisotropic SOC. We note that the analytical
and numerical calculations for the LRO are in close agreement
at low temperatures, but inconsistent at high temperatures where
Bogoliubov theory is expected to be inapplicable. In Figs.~\ref{Fig1}(f)
and~\ref{Fig3}(d), the analytical calculation predict a nonzero
value whereas the numerical one gives a zero value. This discrepancy
is attributed to the fact that Bogoliubov theory is poorly justified
outside the perturbative low temperature regime. 

We have shown that LRO does exist in the relative-phase
sector. But would it imply the existence of an otherwise different
form of BEC? To address this problem, we inspect the single-particle
density matrix (SPDM) for the two-component system defined in analogy
with the scalar BEC {(}see Appendix{)}. Retaining the phase fluctuations,
the matrix elements of the generalized SPDM can be presented as a
2-by-2 matrix: 
\begin{eqnarray}
\mathbf{\rho}(\mathbf{r}',\,\mathbf{r}'') & = & n\left[\begin{array}{cc}
e^{-\frac{\left\langle (\Delta\phi_{\mathrm{t}})^{2}\right\rangle }{2}-\frac{\left\langle (\Delta\phi_{\mathrm{r}})^{2}\right\rangle }{2}} & e^{-\frac{\left\langle (\Delta\phi_{\mathrm{t}})^{2}\right\rangle }{2}-\frac{\left\langle (\Delta_{+}\phi_{\mathrm{r}})^{2}\right\rangle }{2}}\\
e^{-\frac{\left\langle (\Delta\phi_{\mathrm{t}})^{2}\right\rangle }{2}-\frac{\left\langle (\Delta_{+}\phi_{\mathrm{r}})^{2}\right\rangle }{2}} & e^{-\frac{\left\langle (\Delta\phi_{\mathrm{t}})^{2}\right\rangle }{2}-\frac{\left\langle (\Delta\phi_{\mathrm{r}})^{2}\right\rangle }{2}}
\end{array}\right],\label{eq:density-matrix-2}
\end{eqnarray}
where $\Delta_{+}\phi_{\mathrm{r}}=\hat{\phi}_{\mathrm{r}}(\mathbf{r}')+\hat{\phi}_{\mathrm{r}}(\mathbf{r}'')$
{(}see Appendix{)}. The matrix elements of Eq.~\eqref{eq:density-matrix-2}
represent various correlations between atomic fields at different
locations, where the diagonal elements denote the prototypal SPDMs
corresponding to component 1 and 2 respectively. Note that all matrix
elements in Eq.~\eqref{eq:density-matrix-2} contain the prefactor
$e^{-\left\langle (\Delta\phi_{\mathrm{t}})^{2}/2\right\rangle }$,
which vanishes at large distances. As a result, the off-diagonal long-range
order does not extend to the matrix elements of the SPDM implying
that there is no macroscopic eigenvalue and hence the 2D spin-orbit
coupled Bose gas does not exhibit BEC, according to a well-known criterion
for BEC \cite{Leggett2006}. 

As the orientation of local spin density $\mathbf{S}$ is determined
by the relative phase between the components of $\mathbf{\hat{\Psi}}$,
the LRO discussed above manifests a \char`\"{}spin-spin''
correlation. As far as the PW phase is concerned, an anisotropic SOC
is bound to result in two degenerate lowest energy states characterized
by two counter-oriented planar spins, $\pm\mathbf{S}^{0}$, respectively.
This configuration features a 2D Ising-type ground state in the relative-phase
sector and is protected by the energy gap in the dispersion, $\omega_{\mathrm{r}}^{\mathbf{q}}$.
It exhibits LRO by spontaneously breaking $\mathrm{Z}_{2}$
symmetry at finite temperatures. Our numerical simulations suggests
that the Ising-type and BKT transitions occur at the same temperature
$T_{\mathrm{BKT}}^{\infty}$, and the system simultaneously builds
up the quasi and true LROs in the total- and relative-phase
sectors, correspondingly when $T<T_{\mathrm{BKT}}^{\infty}$. It is
interesting to point out that a similar Ising-type phase transition
was predicted to arise in the 2D polar spin-1 condensate subject to
finite quadratic Zeeman energy~\cite{James2011}. For isotropic SOC,
the 2D spin-$\frac{1}{2}$ Bose gas was shown to undergo the BKT transition
at $T=0$~\cite{Jian2011,Liao2014}, suggesting that at any nonzero
temperatures the LRO in the relative-phase sector would
be destroyed by the substantially intensified fluctuations due to
the infinitely degenerate ground state.

\section{conclusions}
We theoretically approach the problem of BEC phase-transition in a 2D Bose gas subjected to anisotropic SOCs. By pinpointing the behavior of correlation functions, we verify that the total phase undergoes a conventional BKT transition, characterized by quasi LRO, whereas the relative phase undergoes an Ising-type transition establishing true LRO. 

It should be noted that we have used a generic SOC Hamiltonian in our model rather than the experimentally realized                      one which has different orientation to the spin-quantization axis and contains extra Rabi and Zeeman terms~\cite{Lin2011}. Nonetheless, 2D generalization of the spin-orbit coupled Bose gas in Ref.~\cite{Lin2011} is shown to possess excitation spectra similar to those in our case~\cite{Martone2012}, and this promises to observe the described phenomena in our study.

Finally, we briefly account for the idea of probing the hidden LRO in the relative phase by means of atomic interferometry.
After the optical pumping by a $\pi/2$ pulse \cite{Egorov2011,Opanchuk2013}, the resultant density of each component
becomes 
\begin{eqnarray}
n_{\pm} & = & \frac{1}{2}\left(\Psi_{1}\pm\Psi_{2}\right)^{*}(\Psi_{1}\pm\Psi_{2}).\label{eq:interference}
\end{eqnarray}
The relative phase $\phi_{r}$ can be extracted from the density fringes
$f$ which is expressed in terms of $n_{\pm}$, 
\begin{eqnarray}
j & = & \frac{n_{+}-n_{-}}{2(n_{+}+n_{-})}\approx\frac{1}{2}\cos\phi_{\mathrm{r}}.\label{eq:fringes}
\end{eqnarray}
The information of the relative phase can be measured through $j$
and the relative-phase correlation can be evaluated.

\begin{acknowledgments}
We thank Blair Blakie, C.-Y. Mou and Ian Spielman for useful discussions. JB thanks
the National Center for Theoretical Sciences at NTHU, Taiwan for hospitality.
SWS, IKL, and SCG are supported by the Ministry of Science and Technology,
Taiwan (Grant No.\ MOST 103-2112-M-018-002-MY3). OF acknowledges
funding from the Dodd-Walls Centre through the New Ideas Fund. This
work was partially supported by the Marsden Fund of New Zealand (Grant
No. MAU1604). 
\end{acknowledgments}

\setcounter{equation}{0} \global\long\def\theequation{A\arabic{equation}}
\setcounter{figure}{0} \global\long\def\thefigure{A\arabic{figure}}

\section*{APPENDIX }

\textbf{\textit{Bogoliubov-de Gennes equation}}. In the the mean-field
approximation, the energy functional of the
spin-orbit coupled 2D Bose gas is 
\begin{eqnarray}
E[\mathbf{\Psi^{*}},\mathbf{\Psi}] & = & \int[\mathbf{\mathbf{\Psi^{*}}}(-\frac{\hbar^{2}}{2m}\nabla^{2}+\kappa_{x}\hat{p}_{x}\hat{\sigma}_{x}+\kappa_{y}\hat{p}_{y}\hat{\sigma}_{y})\mathbf{\Psi}\nonumber \\
 &  & +\frac{g_{11}}{2}\left|\Psi_{1}\right|^{4}+\frac{g_{22}}{2}\left|\Psi_{2}\right|^{4}+g_{12}\left|\Psi_{1}\right|^{2}\left|\psi_{2}\right|^{2}]d^{2}r,\nonumber \\
\label{eq:Energy-SOC}
\end{eqnarray}
where $\mathbf{\Psi}=(\Psi_{1},\Psi_{2})^{\mathrm{T}}$, $\kappa_{x,y}$
the strengths of SOC, $\hat{\sigma}_{x,y}$ the
Pauli matrices and $g_{ij}$ are the nonlinear interaction strengths.
In the following, we consider the case $g_{11}=g_{22}\equiv g$ .
The dynamics is described by the GP equation which can be derived
via the Hartree variational principle $i\hbar\partial_{t}\Psi_{j}=\delta E/\delta\Psi_{j}^{*}=\mathcal{L}_{j}\Psi_{j}$
with $\mathcal{L}_{j}$ the GP evolution operator which takes the
form 
\begin{eqnarray}
i\hbar\partial_{t}\Psi_{1} & = & \left(\frac{-\hbar^{2}}{2m}\nabla^{2}+g\rho_{1}+g_{12}\rho_{2}\right)\Psi_{1}+(\frac{\hbar}{i}\kappa_{x}\partial_{x}-\hbar\kappa_{y}\partial_{y})\Psi_{2},\nonumber \\
i\hbar\partial_{t}\Psi_{2} & = & \left(\frac{-\hbar^{2}}{2m}\nabla^{2}+g\rho_{2}+g_{12}\rho_{1}\right)\Psi_{2}+(\frac{\hbar}{i}\kappa_{x}\partial_{x}+\hbar\kappa_{y}\partial_{y})\Psi_{1},\nonumber \\
\label{eq:GPE}
\end{eqnarray}
where $\rho_{j}=\left|\Psi_{j}\right|^{2}$ is the density of $j$-th
component. In the following calculation, we shall work in the dimessionless
units that the length, time, and energy are scaled by $a_{h}=\sqrt{\hbar/m\omega_{0}}$,
$1/\omega_{0}$, and $\hbar\omega_{0}$ respectively with $m$ the
atomic mass and $\omega$ the transverse trapping frequency. In the
following, the dimensionless interatomic interaction strengths and
SOC strengths are denoted by $\tilde{g}_{ij}$ and $\tilde{\kappa}_{x,y}$,
respectively.

For $g>g_{12}$, the ground state is single plane-wave state (PW)
while for $g<g_{12}$ the ground state is the standing-wave state
which is the superposition of two plane waves carrying two opposite
momenta \cite{Wang2010}. Here we focus on the PW state only that
the ground-state wave function is $\mathbf{\mathbf{\Psi}}^{0}=(\Psi_{1}^{0},\Psi_{2}^{0})^{\mathrm{T}}=\sqrt{n}e^{-i\tilde{\kappa}_{x}x}(1,1)^{\mathrm{T}}$
where we assume the condensation at $\mathbf{p}=(-|\tilde{\kappa}_{x}|,0)$.
To investigate the low-lying excitations, we adopt the Bogoliubov
formulation that the total wave function is decomposed as $\Psi_{j}=e^{-i\mu t}e^{-i\tilde{\kappa}_{x}x}(\Psi_{j}^{0}+\delta\Psi_{j})$
with $\mu$ the chemical potential and $\delta\Psi_{j}$ the low-lying
excitation. We substitute the Bogoliubov decomposition into Eq.~(\ref{eq:GPE})
and retain the correction up to the first order. As a result, the chemical potential
is determined by the zeroth order equation 
\begin{eqnarray}
\mu & = & \left(\tilde{g}+\tilde{g}_{12}\right)n-\frac{\tilde{\kappa}_{x}^{2}}{2},\label{eq:chemical_potential}
\end{eqnarray}
and the first order equation takes the form 
\begin{eqnarray}
i\partial_{t}\delta\Psi_{1} & = & (\frac{-\nabla^{2}}{2}+i\tilde{\kappa}_{x}\partial_{x}+2\tilde{g}n+\tilde{g}_{12}n+\frac{\tilde{\kappa}_{x}^{2}}{2}-\mu)\delta\Psi_{1}+\tilde{g}n\delta\Psi_{1}^{*}\nonumber \\
 &  & +\tilde{g}_{12}n\delta\Psi_{2}+\tilde{g}_{12}n\delta\Psi_{2}^{*}+(\frac{\tilde{\kappa}_{x}}{i}\partial_{x}-\tilde{\kappa}_{y}\partial_{y})\delta\Psi_{2}-\tilde{\kappa}_{x}^{2}\delta\Psi_{2},\nonumber \\
i\partial_{t}\delta\Psi_{2} & = & (\frac{-\nabla^{2}}{2}+i\tilde{\kappa}_{x}\partial_{x}+2\tilde{g}n+\tilde{g}_{12}n+\frac{\tilde{\kappa}_{x}^{2}}{2}-\mu)\delta\Psi_{2}+\tilde{g}n\delta\Psi_{2}^{*}\nonumber \\
 &  & +\tilde{g}_{12}n\delta\Psi_{1}+\tilde{g}_{12}n\delta\Psi_{1}^{*}+(\frac{\tilde{\kappa}_{x}}{i}\partial_{x}+\tilde{\kappa}_{y}\partial_{y})\delta\Psi_{1}-\tilde{\kappa}_{x}^{2}\delta\Psi_{1}.\nonumber \\
\label{eq:first-order}
\end{eqnarray}
Expanding the deviation as $\delta\Psi_{j}=\sum_{\mathbf{q}}(u_{j}^{\mathbf{q}}e^{i(\mathbf{q}\cdot\mathbf{r}-\omega t)}-v_{j}^{\mathbf{q}*}e^{-i(\mathbf{q}\cdot\mathbf{r}-\omega t)})/\sqrt{A}$
with $A$ the area of the system and $\omega$ the excitation energy
of the mode with momentum $\mathbf{q}$ and substituting into Eq.~(\ref{eq:first-order})
yield the Bogoliubov-de Gennes equation 
\begin{widetext}
\begin{equation}
\left(\begin{array}{cccc}
\mathcal{L}_{0}-\tilde{\kappa}_{x}q_{x} & -\tilde{g}n & \tilde{g}_{12}n+h_{soc}-\tilde{\kappa}_{x}^{2} & -\tilde{g}_{12}n\\
\tilde{g}n & -\mathcal{L}_{0}-\tilde{\kappa}_{x}q_{x} & \tilde{g}_{12}n & h_{soc}^{*}+\tilde{\kappa}_{x}^{2}-\tilde{g}_{12}n\\
\tilde{g}_{12}n+h_{soc}^{*}-\tilde{\kappa}_{x}^{2} & -\tilde{g}_{12}n & \mathcal{L}_{0}-\tilde{\kappa}_{x}q_{x} & -\tilde{g}n\\
\tilde{g}_{12}n & h_{soc}+\tilde{\kappa}_{x}^{2}-\tilde{g}_{12}n & \tilde{g}n & -\mathcal{L}_{0}-\tilde{\kappa}_{x}q_{x}
\end{array}\right)\left(\begin{array}{c}
u_{1}^{\mathbf{q}}\\
v_{1}^{\mathbf{q}}\\
u_{2}^{\mathbf{q}}\\
v_{2}^{\mathbf{q}}
\end{array}\right)=\omega\left(\begin{array}{c}
u_{1}^{\mathbf{q}}\\
v_{1}^{\mathbf{q}}\\
u_{2}^{\mathbf{q}}\\
v_{2}^{\mathbf{q}}
\end{array}\right),\label{eq:BdG-1}
\end{equation}
\end{widetext}
where $\mathcal{L}_{0}=q^{2}/2+\tilde{g}n+\tilde{\kappa}_{x}^{2}$,
$h_{soc}=\tilde{\kappa}_{x}q_{x}-i\tilde{\kappa}_{y}q_{y}$ and $u_{j}^{\mathbf{q}},v_{j}^{\mathbf{q}}$
satisfy the normalization condition $\sum_{j}|u_{j}^{\mathbf{q}}|^{2}-|v_{j}^{\mathbf{q}}|^{2}=1$.
For the fully anisotropic SOC ($\tilde{\kappa}_{y}=0$), Eq.~(\ref{eq:BdG-1})
can be diagonalized analytically which yields two distinct dispersion
relations for the excitation modes:
\begin{align}
\omega_{\mathrm{t}}^{\mathbf{q}} & =\sqrt{(\xi_{\mathrm{t}}^{\mathbf{q}})^{2}-(\tilde{g}+\tilde{g}_{12})^{2}n^{2}},\label{eq:excitation-total-1}\\
\omega_{\mathrm{r}}^{\mathbf{q}} & =-2q_{x}\tilde{\kappa}_{x}+\sqrt{(\xi_{\mathrm{r}}^{\mathbf{q}})^{2}-(\tilde{g}-\tilde{g}_{12})^{2}n^{2}}\label{eq:excitation-relative-1}
\end{align}
with the corresponding eigenvectors 
\begin{equation}
\delta\mathbf{\mathbf{\Psi}}_{\mathrm{t}}^{\mathbf{q}}=\frac{1}{2}\left(\begin{array}{c}
\sqrt{\frac{\xi_{\mathrm{t}}^{\mathbf{q}}}{\omega_{\mathrm{t}}^{\mathbf{q}}}+1}\\
\sqrt{\frac{\xi_{\mathrm{t}}^{\mathbf{q}}}{\omega_{\mathrm{t}}^{\mathbf{q}}}-1}\\
\sqrt{\frac{\xi_{\mathrm{t}}^{\mathbf{q}}}{\omega_{\mathrm{t}}^{\mathbf{q}}}+1}\\
\sqrt{\frac{\xi_{\mathrm{t}}^{\mathbf{q}}}{\omega_{\mathrm{t}}^{\mathbf{q}}}-1}
\end{array}\right),\,\delta\mathbf{\mathbf{\Psi}}_{\mathrm{r}}^{\mathbf{q}}=\frac{1}{2}\left(\begin{array}{c}
\sqrt{\frac{\xi_{\mathrm{r}}^{\mathbf{q}}+\omega_{\mathrm{r}}^{\mathbf{q}}+2q_{x}\tilde{\kappa}_{x}}{\omega_{\mathrm{r}}^{\mathbf{q}}+2q_{x}\tilde{\kappa}_{x}}}\\
\sqrt{\frac{\xi_{\mathrm{r}}^{\mathbf{q}}-\omega_{\mathrm{r}}^{\mathbf{q}}-2q_{x}\tilde{\kappa}_{x}}{\omega_{\mathrm{r}}^{\mathbf{q}}+2q_{x}\tilde{\kappa}_{x}}}\\
-\sqrt{\frac{\xi_{\mathrm{r}}^{\mathbf{q}}+\omega_{\mathrm{r}}^{\mathbf{q}}+2q_{x}\tilde{\kappa}_{x}}{\omega_{\mathrm{r}}^{\mathbf{q}}+2q_{x}\tilde{\kappa}_{x}}}\\
-\sqrt{\frac{\xi_{\mathrm{r}}^{\mathbf{q}}-\omega_{\mathrm{r}}^{\mathbf{q}}-2q_{x}\tilde{\kappa}_{x}}{\omega_{\mathrm{r}}^{\mathbf{q}}+2q_{x}\tilde{\kappa}_{x}}}
\end{array}\right),\label{eq:eigenvector-BdG}
\end{equation}
where $\xi_{\mathrm{t}}^{\mathbf{q}}=q^{2}/2+(\tilde{g}+\tilde{g}_{12})n$
and $\xi_{\mathrm{r}}^{\mathbf{q}}=q^{2}/2+(\tilde{g}-\tilde{g}_{12})n+2\tilde{\kappa}_{x}^{2}$.
The bosonic field can be expressed in the form \cite{Pethick2008}
\begin{eqnarray}
\mathbf{\hat{\Psi}}(\mathbf{r}) & = & \left(\begin{array}{c}
\hat{\Psi}_{1}(\mathbf{r})\\
\hat{\Psi}_{2}(\mathbf{r})
\end{array}\right)=e^{i\hat{\phi}_{\mathrm{t}}(\mathbf{r})}\left(\begin{array}{c}
\sqrt{n+\delta n_{1}(\mathbf{r})}e^{i\hat{\phi}_{\mathrm{r}}(\mathbf{r})}\\
\sqrt{n+\delta n_{2}(\mathbf{r})}e^{-i\hat{\phi}_{\mathrm{r}}(\mathbf{r})}
\end{array}\right),\label{eq:fluctuations-1}
\end{eqnarray}
where $\hat{\phi}_{\mathrm{t}}(\mathbf{r})$ and $\hat{\phi}_{\mathrm{r}}(\mathbf{r})$
are respectively the total and relative phase fluctuations. Therefore
for small fluctuations we have {[}expanding Eq.~(\ref{eq:fluctuations-1})
to first order 
\begin{eqnarray}
\hat{\phi}_{\mathrm{t}} & \approx & \frac{1}{4in^{1/2}}\left[(\delta\hat{\Psi}_{1}-\delta\hat{\Psi}_{2}^{\dagger})-h.c.\right],\nonumber \\
\hat{\phi}_{\mathrm{r}} & \approx & \frac{1}{4in^{1/2}}\left[(\delta\hat{\Psi}_{1}-\delta\hat{\Psi}_{2})-h.c.\right],\label{eq:phase_op}
\end{eqnarray}
where $\hat{\phi}_{\mathrm{t}}(\mathbf{r})$ and $\hat{\phi}_{\mathrm{r}}(\mathbf{r})$
are hermitian operators. In the linear approximation, the total and
relative phase operators are decoupled and can be respectively expressed
in terms of the excitations $\delta\mathbf{\Psi}_{\mathrm{t}}^{\mathbf{q}}$
and $\delta\mathbf{\Psi}_{\mathrm{r}}^{\mathbf{q}}$, by writing $\delta\mathbf{\hat{\Psi}}_{\mathrm{t}}^{\mathbf{q}}=\left(\delta\hat{\Psi}_{1,\,\mathrm{t}}^{\mathbf{q}},\,\delta\hat{\Psi}_{2,\,\mathrm{t}}^{\mathbf{q}}\right)^{\mathrm{T}}$
and $\ \delta\mathbf{\hat{\Psi}}_{\mathrm{r}}^{\mathbf{q}}=\left(\delta\hat{\Psi}_{1,\,\mathrm{r}}^{\mathbf{q}},\,\delta\hat{\Psi}_{2,\,\mathrm{r}}^{\mathbf{q}}\right)^{\mathrm{T}}$.
Express the fluctuations as 
\begin{eqnarray}
\delta\hat{\Psi}_{1,\,\mathrm{t}}(\mathbf{r}) & =\delta\hat{\Psi}_{2,\,\mathrm{t}}(\mathbf{r}) & =\frac{1}{\sqrt{A}}\sum_{\mathbf{q}}[u_{\mathrm{t}}^{\mathbf{q}}e^{i\mathbf{q}\cdot\mathbf{r}}\hat{\alpha}_{\mathrm{t}}^{\mathbf{q}}-v_{\mathrm{t}}^{\mathbf{q}*}e^{-i\mathbf{q}\cdot\mathbf{r}}\hat{\alpha}_{\mathrm{t}}^{\mathbf{q}}{}^{\dagger}]\nonumber \\
 &  & =\sum_{\mathbf{q}}[\mathcal{U}_{\mathrm{t}}^{\mathbf{q}}\left(\mathbf{r}\right)\hat{\alpha}_{\mathrm{t}}^{\mathbf{q}}-\mathcal{V}_{\mathrm{t}}^{\mathbf{q}^{*}}(\mathbf{r})\hat{\alpha}_{\mathrm{t}}^{\mathbf{q}}{}^{\dagger}],\nonumber \\
\delta\hat{\Psi}_{1,\,\mathrm{r}}(\mathbf{r}) & =-\delta\hat{\Psi}_{2,\,\mathrm{r}}(\mathbf{r}) & =\frac{1}{\sqrt{A}}\sum_{\mathbf{q}}[u_{\mathrm{r}}^{\mathbf{q}}e^{i\mathbf{q}\cdot\mathbf{r}}\hat{\alpha}_{\mathrm{r}}^{\mathbf{q}}-v_{\mathrm{r}}^{\mathbf{q}*}e^{-i\mathbf{q}\cdot\mathbf{r}}\hat{\alpha}_{\mathrm{r}}^{\mathbf{q}}{}^{\dagger}]\nonumber \\
 &  & =\sum_{\mathbf{q}}[\mathcal{U}_{\mathrm{r}}^{\mathbf{q}}\left(\mathbf{r}\right)\hat{\alpha}_{\mathrm{r}}^{\mathbf{q}}-\mathcal{V}_{\mathrm{r}}^{\mathbf{q}*}(\mathbf{r})\hat{\alpha}_{\mathrm{r}}^{\mathbf{q}}{}^{\dagger}],\label{eq:fluctuations-1-1}
\end{eqnarray}
where $(u_{\mathrm{t}}^{\mathbf{q}},v_{\mathrm{t}}^{\mathbf{q}})=(\sqrt{\frac{\xi_{\mathrm{t}}^{\mathbf{q}}}{\omega_{\mathrm{t}}^{\mathbf{q}}}+1},\sqrt{\frac{\xi_{\mathrm{t}}^{\mathbf{q}}}{\omega_{\mathrm{t}}^{\mathbf{q}}}-1})$,
$(u_{\mathrm{r}}^{\mathbf{q}},v_{\mathrm{r}}^{\mathbf{q}})=(\sqrt{\frac{\xi_{\mathrm{r}}^{\mathbf{q}}+\omega_{\mathrm{r}}^{\mathbf{q}}+2q_{x}\tilde{\kappa}_{x}}{\omega_{\mathrm{r}}^{\mathbf{q}}+2q_{x}\tilde{\kappa}_{x}}},\sqrt{\frac{\xi_{\mathrm{r}}^{\mathbf{q}}-\omega_{\mathrm{r}}^{\mathbf{q}}-2q_{x}\tilde{\kappa}_{x}}{\omega_{\mathrm{r}}^{\mathbf{q}}+2q_{x}\tilde{\kappa}_{x}}})$,
$\hat{\alpha}_{\mathrm{t,r}}^{\mathbf{q}}$ ($\hat{\alpha}_{\mathrm{t,r}}^{\mathbf{q}}{}^{\dagger}$)
the annihilation (creation) operator that destroys (creates) the
excitation in the corresponding branches $\omega_{\mathrm{t,r}}^{\mathbf{q}}$,
 and $(\mathcal{U}_{\mathrm{t,r}}^{\mathbf{q}},\mathcal{V}_{\mathrm{t,r}}^{\mathbf{q}})=(u_{\mathrm{t,r}}^{\mathbf{q}},v_{\mathrm{t,r}}^{\mathbf{q}})e^{i\mathbf{q}\cdot\mathbf{r}}/\sqrt{A}$.

\begin{figure}[ht]
\includegraphics[width=0.9\columnwidth]{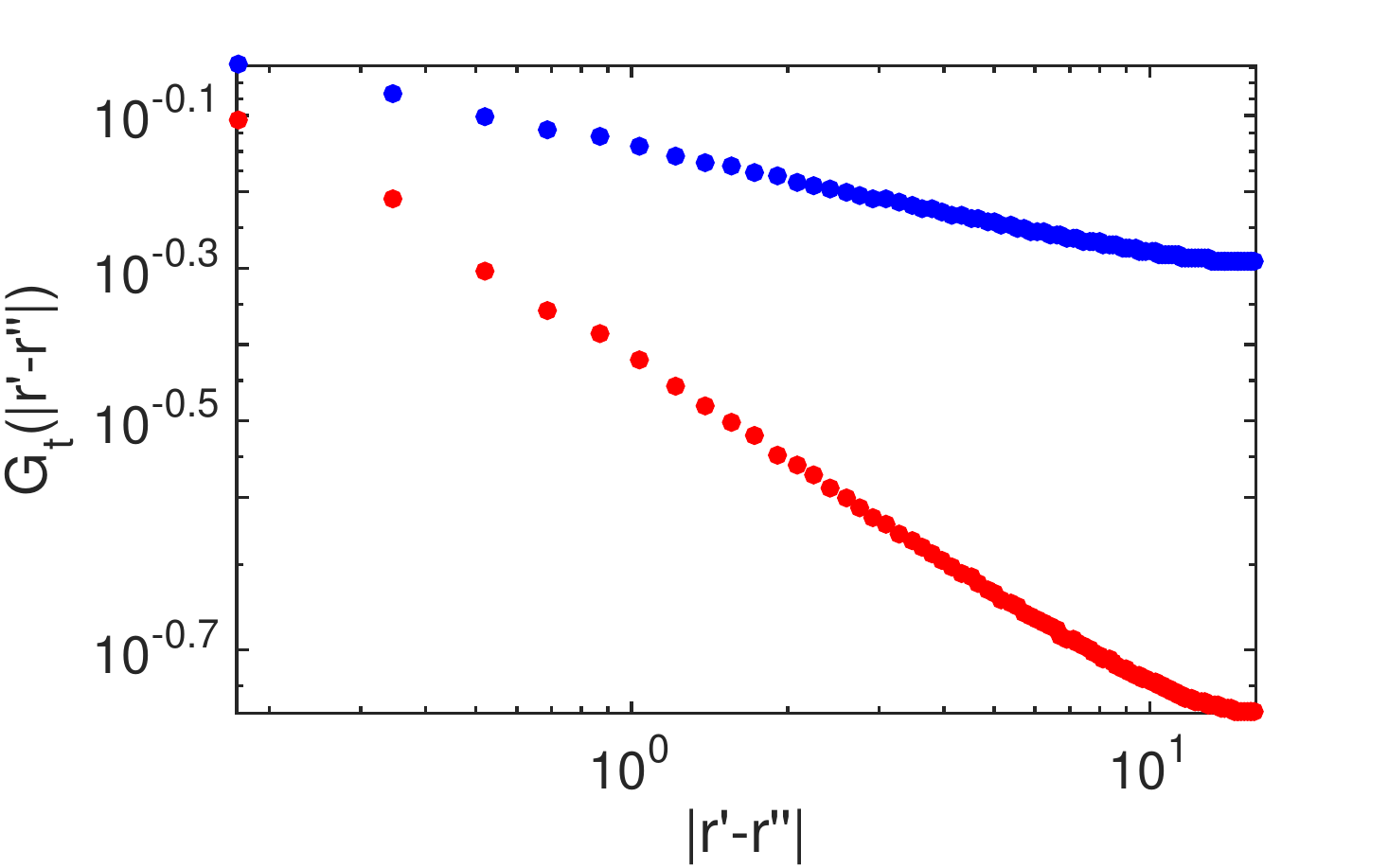}\caption{(Color online). Total-phase correlation functions based on Eq.~(\ref{eq:correlation-1}). Blue and red dots indicate the total-phase correlation functions evaluated at temperatures $T/T_{\mathrm{BKT}}^{\infty}\approx0.44$ and $1.33$ respectively. The linearity of the curves implies a power-law behavior.  }
\label{FigA1} 
\end{figure}

The two-point phase correlation functions are given by 
\begin{equation}
G_{\mathrm{t,r}}\left(\mathbf{r}',\mathbf{r}''\right)=\left\langle e^{i\hat{\phi}_{\mathrm{t,r}}(\mathbf{r}')-i\hat{\phi}_{\mathrm{t,r}}(\mathbf{r''})}\right\rangle =e^{-\left\langle (\Delta\phi_{\mathrm{t,r}})^{2}\right\rangle /2},\label{eq:correlation-1}
\end{equation}
where $\left\langle \cdots\right\rangle $ denotes the ensemble average
and $\Delta\phi_{\mathrm{t,r}}=\hat{\phi}_{\mathrm{t,r}}(\mathbf{r}')-\hat{\phi}_{\mathrm{t,r}}(\mathbf{r}'')$.
The thermal average can be expressed in terms of the Bogoliubov amplitudes
\begin{equation}
\left\langle (\Delta\phi_{\mathrm{t,r}})^{2}\right\rangle =\int\frac{d^{2}q}{\pi n}(N_{\mathrm{t,r}}^{\mathbf{q}}+\frac{1}{2})(u_{\mathrm{t,r}}^{\mathbf{q}}+v_{\mathrm{t,r}}^{\mathbf{q}})^{2}\text{\ensuremath{\sin}}^{2}\frac{\mathbf{q}\cdot\mathbf{r}}{2},\label{eq:fluctuations-int-1}
\end{equation}
where $N_{\mathrm{t,r}}^{\mathbf{q}}=1/(e^{\omega_{\mathrm{t,r}}^{\mathbf{q}}/T}-1)$
is the Bose-Einstein distribution function, $T$ the temperature
measured in units of $\hbar\omega_{0}/k_{B}$, $\mathbf{r}=\mathbf{r}'-\mathbf{r}''$, and
$(u_{\mathrm{t}}^{\mathbf{q}}+v_{\mathrm{t}}^{\mathbf{q}})^{2}=[\xi_{\mathrm{t}}^{\mathbf{q}}+(\tilde{g}+\tilde{g}_{12})n]/2\omega_{\mathrm{t}}^{\mathbf{q}}$
and $(u_{\mathrm{r}}^{\mathbf{q}}+v_{\mathrm{r}}^{\mathbf{q}})^{2}=[\xi_{\mathrm{r}}^{\mathbf{q}}+(\tilde{g}-\tilde{g}_{12})n]/2(\omega_{\mathrm{r}}^{\mathbf{q}}+2\tilde{\kappa}_{x}q_{x})$. Evidently, the total-phase correlation functions so obtained are nothing but precisely the case of a 2D scalar Bose gas, which entails a power-law decay irrespective of temperature~\cite{Pethick2008}, as shown in Fig.~\ref{FigA1} where the functions are plotted on a doubly-logarithmic scale. 

\textbf{\textit{Density matrix.}} For quasi condensates, the density fluctuation is negligible that the density matrix can be expressed as 
\begin{eqnarray}
\mathbf{\rho}(\mathbf{r}',\,\mathbf{r}'') & = & n\left[\begin{array}{cc}
\left\langle e^{i(\phi_{1}(\mathbf{r}')-\phi_{1}(\mathbf{r}''))}\right\rangle  & \left\langle e^{i(\phi_{1}(\mathbf{r}')-\phi_{2}(\mathbf{r}''))}\right\rangle \\
\left\langle e^{i(\phi_{2}(\mathbf{r}')-\phi_{1}(\mathbf{r}''))}\right\rangle  & \left\langle e^{i(\phi_{2}(\mathbf{r}')-\phi_{2}(\mathbf{r}''))}\right\rangle 
\end{array}\right]\label{eq:density-matrix}\\
 & = & n\left[\begin{array}{cc}
e^{-\left\langle \Delta\phi_{1}^{2}\right\rangle /2} & e^{-\left\langle \Delta\phi^{2}\right\rangle /2}\\
e^{-\left\langle \Delta\phi^{2}\right\rangle /2} & e^{-\left\langle \Delta\phi_{2}^{2}\right\rangle /2}
\end{array}\right],\nonumber 
\end{eqnarray}
where $\hat{\phi}_{j}=\hat{\phi}_{\mathrm{t}}+(-1)^{j-1}\hat{\phi}_{\mathrm{r}}$,
$\Delta\phi_{j}=\phi_{j}(\mathbf{r}')-\phi_{j}(\mathbf{r}'')$ and
$\Delta\phi=\phi_{1}(\mathbf{r}')-\phi_{2}(\mathbf{r}'')$. Accordingly, the phase fluctuations are explicitly expressed as

\begin{widetext}
\begin{eqnarray}
(\Delta\phi_{1})^{2} & = & (\Delta\phi_{\mathrm{t}})^{2}+(\Delta\phi_{\mathrm{r}})^{2}+2\left[\hat{\phi}_{\mathrm{t}}(\mathbf{r}')\hat{\phi}_{\mathrm{r}}(\mathbf{r}')+\hat{\phi}_{\mathrm{t}}(\mathbf{r}'')\hat{\phi}_{\mathrm{r}}(\mathbf{r}'')-\hat{\phi}_{\mathrm{t}}(\mathbf{r}')\hat{\phi}_{\mathrm{r}}(\mathbf{r}'')-\hat{\phi}_{\mathrm{t}}(\mathbf{r}'')\hat{\phi}_{\mathrm{r}}(\mathbf{r}')\right],\nonumber \\
(\Delta\phi_{2})^{2} & = & (\Delta\phi_{\mathrm{t}})^{2}+(\Delta\phi_{\mathrm{r}})^{2}-2\left[\hat{\phi}_{\mathrm{t}}(\mathbf{r}')\hat{\phi}_{\mathrm{r}}(\mathbf{r}')+\hat{\phi}_{\mathrm{t}}(\mathbf{r}'')\hat{\phi}_{\mathrm{r}}(\mathbf{r}'')-\hat{\phi}_{\mathrm{t}}(\mathbf{r}')\hat{\phi}_{\mathrm{r}}(\mathbf{r}'')-\hat{\phi}_{\mathrm{t}}(\mathbf{r}'')\hat{\phi}_{\mathrm{r}}(\mathbf{r}')\right],\nonumber \\
(\Delta\phi)^{2} & = & (\Delta\phi_{\mathrm{t}})^{2}+(\hat{\phi}_{\mathrm{r}}(\mathbf{r}')+\hat{\phi}_{\mathrm{r}}(\mathbf{r}''))^{2}+2[\hat{\phi}_{\mathrm{t}}(\mathbf{r}')\hat{\phi}_{\mathrm{r}}(\mathbf{r}')-\hat{\phi}_{\mathrm{t}}(\mathbf{r}'')\hat{\phi}_{\mathrm{r}}(\mathbf{r}'')-\hat{\phi}_{\mathrm{t}}(\mathbf{r}')\hat{\phi}_{\mathrm{r}}(\mathbf{r}'')+\hat{\phi}_{\mathrm{t}}(\mathbf{r}'')\hat{\phi}_{\mathrm{r}}(\mathbf{r}')]\label{eq:variance-1-2}
\end{eqnarray}
\end{widetext}

Equation (\ref{eq:density-matrix}) can be simplified as 
\begin{eqnarray}
\mathbf{\rho}(\mathbf{r}',\,\mathbf{r}'') & = & n\left[\begin{array}{cc}
e^{-\left\langle (\Delta\phi_{\mathrm{t}})^{2}/2\right\rangle }e^{-\left\langle (\Delta\phi_{\mathrm{r}})^{2}/2\right\rangle } & e^{-\left\langle (\Delta\phi_{\mathrm{t}})^{2}/2\right\rangle }e^{-\left\langle (\Delta_{+}\phi_{\mathrm{r}})^{2}/2\right\rangle }\\
e^{-\left\langle (\Delta\phi_{\mathrm{t}})^{2}/2\right\rangle }e^{-\left\langle (\Delta_{+}\phi_{\mathrm{r}})^{2}/2\right\rangle } & e^{-\left\langle (\Delta\phi_{\mathrm{t}})^{2}/2\right\rangle }e^{-\left\langle (\Delta\phi_{\mathrm{r}})^{2}/2\right\rangle }
\end{array}\right],\nonumber \\
\label{eq:density-matrix-2-1}
\end{eqnarray}
where $\Delta_{+}\phi_{\mathrm{r}}=\hat{\phi}_{\mathrm{r}}(\mathbf{r}')+\hat{\phi}_{\mathrm{r}}(\mathbf{r}'')$
and the full density matrix consists of block matrices. The ensemble
average of the cross terms in Eq.~(\ref{eq:variance-1-2}) vanish
since $\left\langle \hat{\alpha}_{\mathrm{t}}^{\mathbf{q}}\hat{\alpha}_{\mathrm{r}}^{\mathbf{q}}\right\rangle =\left\langle \hat{\alpha}_{\mathrm{t}}^{\mathbf{q}\dagger}\hat{\alpha}_{\mathrm{r}}^{\mathbf{q}}\right\rangle =\left\langle \hat{\alpha}_{\mathrm{t}}^{\mathbf{q}\dagger}\hat{\alpha}_{\mathrm{r}}^{\mathbf{q}\dagger}\right\rangle =\left\langle \hat{\alpha}_{\mathrm{t}}^{\mathbf{q}}\hat{\alpha}_{\mathrm{r}}^{\mathbf{q}\dagger}\right\rangle =0$.
Following the same procedure of deriving Eq.~(\ref{eq:fluctuations-int-1}),
the expectation value of $(\Delta_{+}\phi_{\mathrm{r}})^{2}$ is 
\begin{eqnarray}
\left\langle (\Delta_{+}\phi_{\mathrm{r}})^{2}\right\rangle  & = & \int\frac{d^{2}q}{\pi n}\,(N_{\mathrm{r}}^{\mathbf{q}}+\frac{1}{2})(u_{\mathrm{r}}^{\mathbf{q}}+v_{\mathrm{r}}^{\mathbf{q}})^{2}\cos^{2}\frac{\mathbf{q}\cdot\mathbf{r}}{2},\nonumber \\
\label{eq:phi_r-+}
\end{eqnarray}
where $\left\langle (\Delta_{+}\phi_{\mathrm{r}})^{2}\right\rangle $
is maximized at $\mathbf{r}=0$. Therefore in the thermodynamics limit,
the off-diagonal elements of the block matrices vanishes when $|\mathbf{r}'-\mathbf{r}''|\rightarrow\infty$
since all the block matrices contain the same prefactor $e^{-\left\langle (\Delta\phi_{\mathrm{t}})^{2}/2\right\rangle }$
which would destroy the off-diagonal LRO. The absence
of off-diagonal LRO of the full density matrix Eq.~(\ref{eq:density-matrix})
indicates that the 2D SO coupled Bose gas could not undergo the Bose-Einstein
condensation even though the relative-phase sector could possess a
true long-range order.

\bibliographystyle{apsrev4-1}
%

\end{document}